\documentclass[10pt,conference]{IEEEtran} 

\usepackage{float}
\usepackage[utf8]{inputenc} 
\usepackage[T1]{fontenc}    
\usepackage[hidelinks,pdftex,
            pdfauthor={Benjamin Steenhoek, Michele Tufano, Neel Sundaresan, and Alexey Svyatkovskiy},
            pdftitle={Reinforcement Learning from Automatic Feedback for High-Quality Unit Test Generation},
            pdfkeywords={deep learning, test generation, reinforcement learning, LLM}]{hyperref}       

\usepackage{url}            
\usepackage{booktabs}       
\usepackage{amsfonts}       
\usepackage{nicefrac}       
\usepackage{microtype}      
\usepackage{tcolorbox}
\usepackage{adjustbox}
\usepackage{multirow}
\usepackage{graphicx}
\usepackage{subcaption}
\definecolor{bg}{HTML}{282828}
\usepackage{pmboxdraw}
\usepackage{algorithmic}
\usepackage{listings}
\usepackage[linesnumbered,ruled,vlined]{algorithm2e}
\def\BibTeX{{\rm B\kern-.05em{\sc i\kern-.025em b}\kern-.08em
    T\kern-.1667em\lower.7ex\hbox{E}\kern-.125emX}}

\usepackage{enumitem}
\usepackage{amsmath}
\usepackage{flushend}
\usepackage{tikz}

\usepackage{pifont}
\newcommand{\cmark}{\ding{51}}%
\newcommand{\xmark}{\ding{55}}%

\usepackage[square,numbers]{natbib}

\usepackage{makecell}
\usepackage{xcolor}

\usepackage{tikzducks}

\usetikzlibrary{decorations.pathreplacing,calc,tikzmark}

\usepackage{graphicx}
\usepackage[super]{nth}
\usepackage{MnSymbol}

\usepackage{xfrac}

\usepackage[symbol]{footmisc}

\usepackage{tablefootnote}

\usepackage{mdframed}

\usepackage{listings}
\usepackage{color}
\usepackage{tikz}
\usetikzlibrary{arrows.meta, positioning}
\usetikzlibrary{calc}

\usepackage[frozencache,cachedir=.]{minted}
\usepackage{tabularx}
\usepackage{tabularray}

\usepackage{dblfloatfix}
\usepackage{mathtools}
\usepackage{tablefootnote}

\usepackage[nameinlink]{cleveref}

\newcommand{\nb}[2]{
    \fbox{\bfseries\sffamily\scriptsize#1}
    {\sf\small$\blacktriangleright$\textit{#2}$\blacktriangleleft$}
}

\newcommand\ben[1]{\textcolor{red}{BEN: [{#1}]}}
\newcommand\temp[1]{#1}

\newcommand\ALEXEY[1]{\textcolor{green}{\nb{ALEXEY}{#1}}}

\SetCommentSty{mycommfont}

\SetKwInput{KwInput}{Input}                
\SetKwInput{KwOutput}{Output}              

\lstdefinestyle{myJavaStyle}{
  frame=tb,
  float=*,
  language=java,
  aboveskip=3mm,
  belowskip=3mm,
  showstringspaces=false,
  columns=flexible,
  basicstyle={\small\ttfamily},
  numbers=none,
  numberstyle=\tiny\color{gray},
  keywordstyle=\color{blue},
  commentstyle=\color{dkgreen},
  stringstyle=\color{mauve},
  frame=single,
  breaklines=true,
  breakatwhitespace=true,
  tabsize=3,
}

\newcommand{\mydarkred}{black!20!red}
\newcommand{\mydarkgreen}{black!30!green}
\newcommand{\goodmark}{\textcolor{\mydarkgreen}{\cmark}}
\newcommand{\badmark}{\textcolor{\mydarkred}{\xmark}}

\newcommand{\ppomodel}[1]{\textit{#1}}

\newcommand{\figref}[1]{\tikz[baseline={(char.base)}]{
            \node[shape=circle,draw,fill=black,inner sep=0.8pt, text=white] (char) {#1};}}

\setcitestyle{square}

\begin{document}



\title{Reinforcement Learning from Automatic Feedback for High-Quality Unit Test Generation}

\author{
    \IEEEauthorblockN{Benjamin Steenhoek\IEEEauthorrefmark{1}, Michele Tufano\IEEEauthorrefmark{2}, Neel Sundaresan\IEEEauthorrefmark{1}, Alexey Svyatkovskiy\IEEEauthorrefmark{3}}
    \IEEEauthorblockA{\IEEEauthorrefmark{1}Microsoft Data \& AI
    \\\{bensteenhoek,neels\}@microsoft.com}
    \IEEEauthorblockA{\IEEEauthorrefmark{2}Google, \IEEEauthorrefmark{3}Google DeepMind
    \\\{tufanomichele,alexeysv\}@google.com}
}







\maketitle

\begin{abstract}
Software testing is a crucial but time-consuming aspect of software development, and recently, Large Language Models (LLMs) have gained popularity for automated test case generation. However, because LLMs are trained on vast amounts of open-source code, they often generate test cases that do not adhere to best practices and may even contain test smells (anti-patterns).
To address this issue, we propose Reinforcement Learning from Static Quality Metrics (RLSQM), wherein we utilize Reinforcement Learning to generate high-quality unit tests based on static analysis-based quality metrics.
First, we analyzed LLM-generated tests and show that LLMs frequently do generate undesirable test smells --- up to 37\% of the time.
Then, we implemented lightweight static analysis-based reward model and trained LLMs using this reward model to optimize for five code quality metrics.
Our experimental results demonstrate that the RL-optimized Codex model consistently generated higher-quality test cases than the base LLM, improving quality metrics by up to 23\%, and generated nearly 100\% syntactically-correct code. RLSQM also outperformed GPT-4 on all code quality metrics, in spite of training a substantially cheaper Codex model.
We provide insights into how reliably utilize RL to improve test generation quality and show that RLSQM is a significant step towards enhancing the overall efficiency and reliability of automated software testing.
Our data are available at this link: \url{https://doi.org/10.6084/m9.figshare.25983166}.

\end{abstract}

\section{Introduction}



Software testing is a crucial component in the development of reliable and robust software systems. In addition, tests must be maintainable to promote efficient software development and bug detection.
\textit{Test smells} encompass a range of issues that can hinder the comprehensibility, maintainability, and overall quality of a test suite. Numerous studies have highlighted the detrimental impact of test smells on various aspects of software testing~\cite{van2001refactoring,bavota2015test,tufano_empirical_2016,spadini2018relation,peruma_distribution_2019,spadini_investigating_2020}.
In particular, Bavota et al.~\cite{bavota2015test} showed that test smells are correlated with decreased comprehensibility and maintainability of the test code, and Spadini et al.~\cite{spadini2018relation} demonstrated a connection between test smells and increased susceptibility to changes and defects in the production code.
Consequently, test smells must be addressed to ensure the effectiveness of software testing efforts.

Large Language Models (LLMs) have become popular for automatically generating code~\cite{codex,copilot}, including test cases~\cite{codamosa,xie2023chatunitest,yuan2023manual,schäfer2023empirical,dakhel2023effective}.
However, because LLMs are trained on open source code, they often incorporate undesirable properties seen in real-world test suites.
\temp{In code completion, this manifests when LLMs generate vulnerable code~\cite{asleep_at_the_keyboard} or leak sensitive data~\cite{seatbelt,codexleaks}.}
We show in \Cref{sec:rq1-results} that pre-trained models often exhibit test smells when generating tests.

Recently, Reinforcement Learning (RL) has been used to train LLMs to  follow instructions~\cite{instructgpt}, instill specific qualities~\cite{rlaif,helpful,llama2} and generate higher-quality code~\cite{codellama,grammformer}. RL addresses \textit{exposure bias}~\cite{exposure_bias1, exposure_bias2} which arises during supervised fine-tuning (SFT), where models are trained on ground-truth text but rely on self-generated outputs to generate the next word, causing a distribution mismatch. It also avoids the additional inference cost of prompting approaches like in-context learning~\cite{liu_few-shot_2022} and tree-of-thoughts~\cite{yao_tree_2023}.


To address these issues, we introduce \textit{Reinforcement Learning from Static Quality Metrics (RLSQM)}, which uses RL to align LLMs with a static analysis-based reward model.
We leverage the framework of Reinforcement Learning from Human Feedback (RLHF)~\cite{chatgpt,rlhf_2020,instructgpt,reddit_rlhf,christiano2023deep}, which overcomes the data limitations of traditional fine-tuning by training LLMs to align with a reward model.
Instead of relying on expensive, unpredictable, and often biased human feedback, we use fully automated static analysis to detect well-known quality metrics. We train a Reward Model (RM) to score test cases based on these quality metrics, then use it to provide feedback for the Proximal Policy Optimization (PPO)~\cite{ppo} algorithm to train LLMs to generate test cases that maximize the expected reward (i.e., higher quality tests).


We begin by generating test cases using base LLMs and investigating their alignment with testing best practices and their susceptibility to test smells (\S~\ref{sec:rq1-design}). Our results show that LLM-generated test cases often fail to follow best practices contain undesirable test smells; specifically, base Codex generated 17\% tests with incorrect syntax, 31\% tests without assertions, and 37\% tests without calls to the focal method (method under test).

Next, we apply RLSQM and show that it enables LLMs to generate substantially higher-quality tests (\S~\ref{sec:rq2-design}).
We explored several ablations and strategies to optimize models for multiple quality metrics, providing valuable insight into effective RL fine-tuning for test generation.
RL-finetuned Codex models generated more tests with best practices and fewer containing test smells: up to 23.2\% more tests with Assertions and 17.9\% more tests with Focal calls, and up to 2.22\% fewer tests with Duplicate Assertions and 2.53\% fewer tests with Conditionals/Exceptions.
Furthermore, compared to the substantially more expensive GPT-4 model, RL-finetuned models yielded greater performance on five out of seven quality metrics, including all code-quality metrics, at a substantially lower inference cost.

In summary, we make several contributions:
\begin{itemize}
    \item We analyze over 6 million LLM-generated tests with respect to Best Practices, Documentation, and Test Smells.
    \item We introduce Reinforcement Learning from Static Quality Metrics (RLSQM), a fully-automatic approach for fine-tuning language models to generate high-quality tests.
    \item We train and evaluate RLSQM on a dataset of 62 thousand C\# focal methods from 82 open-source projects.
    \item We explore diverse strategies for training with RLSQM, including Individual, Sequential, and Combined Rewards, and report on settings which were critical for convergence.
\end{itemize}


\section{Approach}

\temp{In RLSQM, we follow these steps: we collect open-source methods we want to test (\S~\ref{sec:data-collection}), generate unit tests with base LLMs (\S~\ref{sec:prompting}), automatically evaluate the quality of the tests with Quality Analyzer (\S~\ref{sec:quality-analyzer}), filter to a set of high-quality ``golden'' tests for supervised fine-tuning (\S~\ref{sec:sft}), then use Quality Analyzer directly in RL fine-tuning (\S~\ref{sec:rl}).}

\subsection{Data collection}
\label{sec:data-collection}

\begin{table}[htbp]
    \centering
    \caption{Statistics of the dataset of focal methods.
    }
    \label{tab:dataset-statistics}
    \begin{tabular}{lr}
    \toprule
    \textbf{Statistic} & \textbf{Value} \\
    \midrule
    \# projects                   & 82 \\
    \# focal classes              & 13,567 \\
    \# focal methods              & 62,103 \\
    Mean \# SLOC per focal method       & 6.5 \\
    Mean \# parameters per focal method & 1.6 \\
    \bottomrule
    \end{tabular}
\end{table}

We constructed our dataset by curating focal methods from open-source C\# projects on GitHub. Initially, we identified 100 non-fork projects predominantly written in C\# with a minimum of 5 stars. Among these, we exclusively considered projects featuring TestMethods written in the MSTest framework~\cite{MSTest2023}, ultimately selecting 82 repositories. From these repositories, we extracted all public methods for which the developer wrote tests, in order to focus our study on the methods which developers specifically intended to assess. Our dataset consists of 62k focal methods sourced from these 82 projects, averaging around 757 focal methods per project.
\Cref{tab:dataset-statistics} lists the statistics characterizing our dataset.


\subsection{Generating Unit Tests with Language Models}
\label{sec:prompting}

We formulate the unit test generation task as completing a test case based on a focal method and its context -- the focal class source code and relevant file-level context.
This formulation aligns with the causal language modeling objective employed during pre-training.
\temp{While we employed this prompt design because it consistently performed the best in our pilot studies, we emphasize that our findings are orthogonal to prompt design, and can be integrated with a diverse set of prompts.}


\Cref{fig:test-generation-example} details the prompt's contents (top to bottom):
\begin{itemize}
    \item The filepath containing the focal method.
    \item The focal method's source code and accompanying context.
    \item A ``hint'' specifying a hypothetical filepath for the test code and the beginning of the test method signature, which conditions the model to test the focal method.
\end{itemize}

\begin{figure}[t]
    \centering




\definecolor{color1}{rgb}{0.95,0.95,0.92}
\definecolor{lightgreen}{HTML}{ACE1AF}
\definecolor{lightyellow}{rgb}{0.992, 0.992, 0.588}
\newcommand{\vspaceremove}{8pt}

\usemintedstyle{vs}

\begin{minipage}{\linewidth}
\end{minipage}
\vspace*{-16pt}
\begin{minipage}{\linewidth}
    \begin{minipage}{0.075\linewidth}
        \rotatebox{90}{
        $\overbrace{\qquad\qquad\qquad\qquad\qquad\qquad\qquad\qquad\qquad\qquad\qquad\qquad\qquad\qquad\qquad}^{\text{Prompt}}$}
    \end{minipage}%
    \begin{minipage}{0.85\linewidth}
    \begin{minted}[breaklines,tabsize=2,bgcolor=color1,highlightlines={1,8-12,14-18,20-23},highlightcolor=lightyellow]{cs}
src/Commands/BenchmarkCommand.cs:
using System;
using System.Linq;
using System.Threading.Tasks;

public class BenchmarkCommand : ICommand
{
	private WriteBenchmarkCommand _writeBenchmark = new WriteBenchmarkCommand();
	private ReadBenchmarkCommand _readBenchmark = new ReadBenchmarkCommand();

	// Other Fields
	// Other Methods

	public async Task Stop()
	{
		await _writeBenchmark.Stop();
		await _readBenchmark.Stop();
	}
}
src/Commands/TestBenchmarkCommand.cs:
[TestMethod]
public void TestStop()
    \end{minted}
    \end{minipage}%
    \begin{minipage}{0.075\linewidth}
        \rotatebox{270}{
        $
        \phantom{x}
        \overbrace{\phantom{x}}^{\text{Filepath}}
        \phantom{xxxxxxxxxxxx}
        \overbrace{\phantom{xxxxxxxxxxxxxxxxxx}}^{\text{Adaptive Focal Context}}
        \phantom{xx}
        \overbrace{\phantom{xxxxxxxxxxx}}^{\text{Focal Method}}
        \phantom{xx}
        \overbrace{\phantom{xxxxxx}}^{\text{Hint}}
        \phantom{xx}\;
        $}
    \end{minipage}%
\end{minipage}
\vspace*{-\vspaceremove}
%
%
%
\begin{minipage}{\linewidth}
    \begin{minipage}{0.075\linewidth}
        \rotatebox{90}{
        $\overbrace{\phantom{xxxxxxxxxxxxxxxxxxx}}^{\text{Completion}}$}
    \end{minipage}%
    \begin{minipage}{0.85\linewidth}
    \begin{minted}[breaklines,tabsize=2,bgcolor=color1,highlightlines={1-5},highlightcolor=lightgreen]{cs}
{
	var command = new BenchmarkCommand();
	command.Stop().Wait();
	Assert.IsTrue(command.IsStopped());
}
src/Commands/TestBenchmarkCommand.cs:
[TestMethod]
public void TestBenchmarkCommand()
// More generated text...
    \end{minted}
    \end{minipage}%
    \begin{minipage}{0.075\linewidth}
        \rotatebox{270}{
        $
        \;
        \overbrace{\phantom{xxxxxxxxxx}\;}^{\text{Generated test}}
        \;
        \overbrace{\phantom{xxxxxxxxx}}^{\text{Extra text}}
        $}
    \end{minipage}%
\end{minipage}


    \caption{Our prompt design for test generation. The context given to the model is truncated to fit within the model's context length.}
    \label{fig:test-generation-example}
\end{figure}



We employ \textit{adaptive focal context}~\cite{athenatest} to ensure the inputs fit within the model's context length, trying first the entire focal file, then abbreviating context methods, fields, and comments. To generate diverse test suites, we set $temperature = 0.7$, $top\_p = 1.0$, and $frequency\_penalty = 0.5$ and generated multiple completions for each prompt. For complete details, see \Cref{sec:prompt-details}.

\subsection{Evaluating Test Case quality}
\label{sec:quality-analyzer}

We assessed the properties of the generated unit tests with respect to commonly-accepted best practices (e.g. syntactic correctness) and test-smells (e.g. missing assert statement, missing call to the focal method, or conditional logic); see \Cref{sec:rq1-design} for the full list of metrics.
We implemented a tool which we call \textit{Quality Analyzer} to evaluate these quality metrics. Inspired by tsDetect for Java~\cite{tsdetect}, our tool uses \textit{tree-sitter}~\cite{treesitter} to traverse the Abstract Syntax Tree (AST) and analyze the code's quality properties.
We used Quality Analyzer to automatically provide the feedback needed to construct a dataset for SFT and perform RL training.
This tool is available in our data package~\cite{data-package}.

\subsection{Supervised Fine-tuning}\label{sec:sft}

\begin{figure}[tb]
    \centering
    \includegraphics[width=1.0\linewidth]{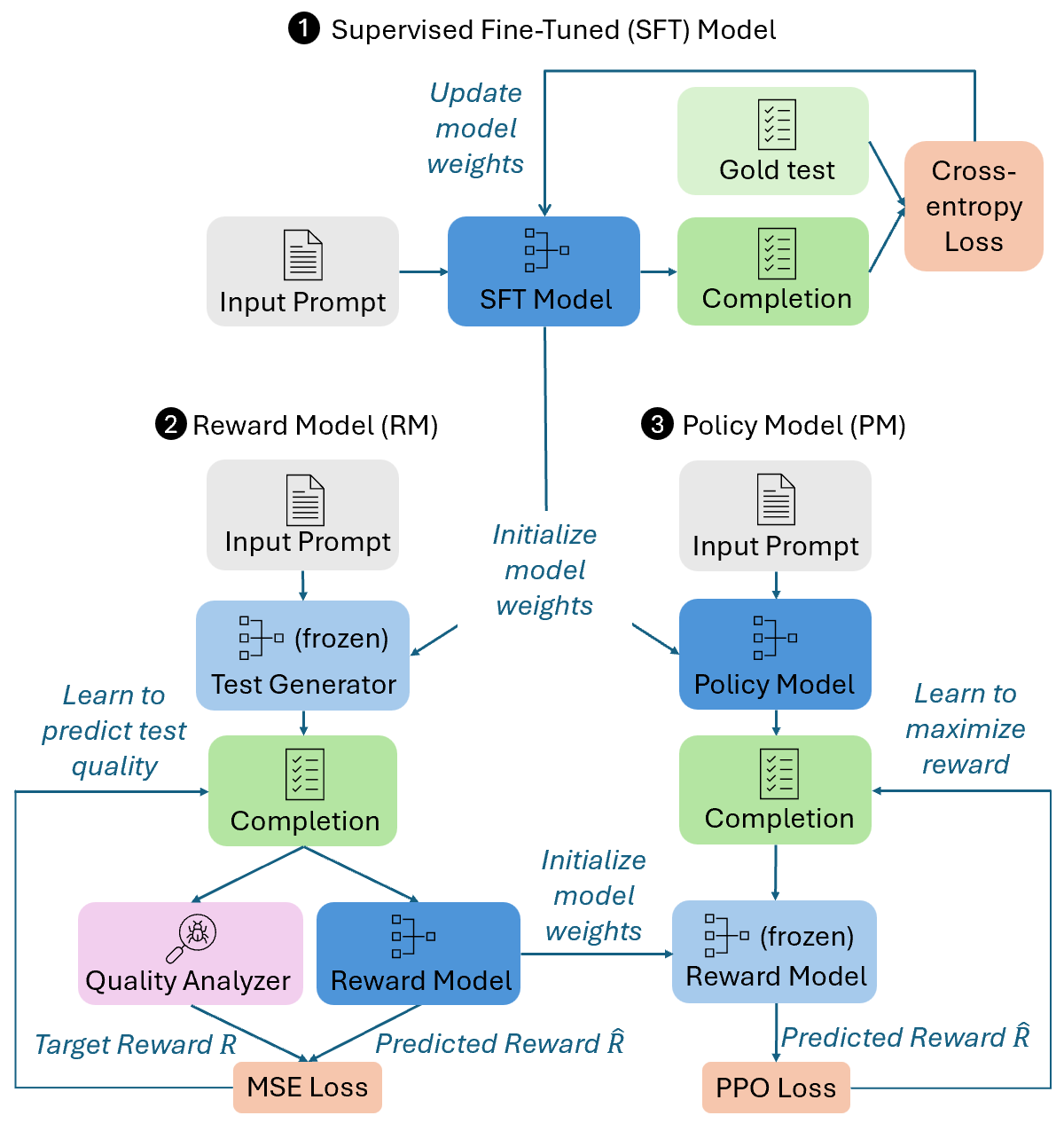}

    \caption{Overview of our approach: Reinforcement Learning from Static Quality Metrics (RLSQM).
    (1) We automatically generate tests and filter them to a \textit{golden set} of high-quality of using Quality Analyzer, then train the model to generate high-quality tests with supervised fine-tuning (SFT).
    (2) We generate tests with the SFT model and use Quality Analyzer to train the reward model (RM) to predict accurate reward scores.
    (3) We initialize the policy model (PM) with the SFT model weights and fine-tune it with PPO to maximize the rewards predicted by the RM.
    }
    \label{fig:sft}
\end{figure}

\Cref{fig:sft}~(\figref{1}) illustrates the supervised fine-tuning (SFT) process.
We used the Quality Analyzer to create a \textit{golden dataset} for supervised fine-tuning.
We filtered the unit test to exhibit syntactic correctness, contain at least one assertion, and invoke the focal method, while avoiding duplicate assertions and conditionals.
%
%
%
We then fine-tune the language model to generate these high-quality tests by minimizing the cross-entropy loss.

By filtering the dataset to only include test cases with desirable qualities, we are effectively training a \textit{refined} version of the model which generates higher-quality test cases.
\temp{Although this approach consistently produced higher-quality test cases, our experimental results (discussed in \Cref{sec:experiments}) highlight some limitations of this approach, including the fact that it cannot learn from negative examples.
In contrast, RL enables the model to actively penalize undesirable behaviors and test smells, offering a more robust training process.}

\subsection{Reinforcement Learning from Static Quality Metrics}
\label{sec:rl}

We frame reinforcement learning fine-tuning as an episodic RL scenario~\cite{sutton2018reinforcement}.
In each episode, the agent generates a unit test case by sequencing tokens from the vocabulary of tokens.
Each action generates a single token, each observation consists of the prompt + the sequence of tokens generated in previous steps, and the episode ends when the model generates a \textit{stop} token or reaches the maximum number of tokens.
The agent relies on a policy $\pi$, instantiated by a pretrained language model, to decide its next action. This policy seeks to guide the agent in making decisions that maximize the episodic reward. 
At the end of the episode, the agent receives a reward $\mathcal{R}$ which corresponds to the successful completion of a unit test case.

\subsubsection{Reward Model Training}



\Cref{fig:sft}~(\figref{2}) zooms in on the RLSQM reward model (RM) finetuning stage.
As in RLHF, the reward model is pivotal to our approach. However, rather than relying on human feedback to train the RM, we employ a static Quality Analyzer to automatically determine a scalar reward score for each prompt and completion (see \S~\ref{sec:training-strategies} for details of the reward score). These derived rewards are then harnessed to train a model to estimate expected rewards from the agent's decisions. 

The RM is trained by optimizing the Mean Squared Error (MSE) loss, given by:
\begin{equation}
\text{MSE} = \frac{1}{n} \sum_{i=1}^{n} (R(x_i, y_i) - \hat{R}(x_i, y_i))^2
\end{equation}
In this equation, $R(x_i, y_i)$ is the reward for one generated test $y_i$ with prompt $x_i$ as per the Quality Analyzer, and $\hat{R}(x_i, y_i)$ is its predicted counterpart.

The RM learns to predict a scalar reward score that closely matches the Quality Analyzer.
%
While the Quality Analyzer could directly provide scores suitable for RL fine-tuning, we note that the RM is more general and can easily extend to less-defined rewards, including human feedback; it can learn to combine multiple quality metrics; and it can estimate a reward score even when static analysis tools fail (incorrect syntax or partial code snippets).

\subsubsection{PPO Fine-tuning}

\Cref{fig:sft}~(\figref{3}) displays the RLSQM's policy model (PM) finetuning stage.
RLSQM utilizes the Proximal Policy Optimization (PPO)~\cite{ppo} algorithm to train the PM. We initialized the policy $\pi$ using the SFT model and initialized the value function using the weights of the reward model (following~\cite{instructgpt}).
The PM learns to optimize the reward, minus a penalty when it generates tokens that are dramatically different from the base model (also following \cite{instructgpt}):



\begin{equation}
    \mathcal{R}(x, y) = \left[ \hat{R}(x, y) - \beta D_{\text{KL}}(\pi_{\theta_{0}}, \pi_{\theta}) \right]
\end{equation}

Here, $\pi_{\theta_0}$ represents the initial weights of the policy before PPO fine-tuning, $\pi_{\theta}$ represents the policy's current weights, $D_{\text{KL}}$ denotes the function computing the KL divergence, and and $\beta$ is a hyperparameter scaling the KL divergence penalty.

The policy model is trained to maximize $\mathcal{R}(x, y)$ by minimizing the clipped surrogate objective:

\begin{equation}
L^{\text{CLIP}}(\theta) = \mathbb{E}\left[ \min \left( r_t(\theta) A^{\text{old}}, \text{clip}_{1\pm\epsilon}\left(r_t(\theta)\right) A^{\text{old}} \right) \right]
\end{equation}

Where the ratio \( r_t(\theta) = \frac{\pi_\theta(a_t|s_t)}{\pi_{\theta_{\text{old}}}(a_t|s_t)} \) contrasts the new policy's probability to the old's for action \(a_t\) in state \(s_t\), \( A^{\text{old}} \) is the advantage estimate on state-action pair $(s_t, a_t)$ from the old policy, which is computed based on $\mathcal{R}$, and $\epsilon$ is a PPO clipping parameter that confines the policy's update magnitude.

\temp{Our approach is not tied to the PPO algorithm; implementations may employ a variety of other RL algorithms, such as A2C~\cite{mnih2016asynchronous} or NLPO~\cite{ramamurthy2023reinforcement}.}



\subsection{Training strategies for RLSQM}
\label{sec:training-strategies}

Our approach of training the reward model is general and therefore, many possible strategies exist for selecting the target reward function ($R$). We propose three training strategies which assign negative reward to discourage syntactically-incorrect completions and grant positive rewards based on quality properties.


\noindent\textbf{Individual Reward:}
In the most basic setting, the target reward function, denoted $R_P(x_i,y_i)$ for some prompt and completion $x_i, y_i$ and property $P$, is assigned based on whether the test is syntactically correct and exhibits $P$, according to \Cref{tab:reward-tables} (left).
The resulting reward model provides higher rewards to test cases that adhere to the chosen quality metric during the PPO fine-tuning phase.


\begin{table}[h]
    \centering
    \caption{Target reward ($R$): Individual Reward for a property $P$ (left) and Combined Reward for 2 properties $P$ and $Q$ (right).}
    \label{tab:reward-tables}


    \begin{tabular}[b]{ccrcccr}
    \multicolumn{3}{c}{\textbf{Individual Reward}} & \multicolumn{4}{c}{\textbf{Combined Reward}} \\
    \cmidrule(r){1-3} \cmidrule(l){4-7}
    \makecell[l]{Correct\\Syntax} & $P$ & Reward & \makecell[l]{Correct\\Syntax} & $P$ & $Q$ & Reward \\
    \cmidrule(r){1-3} \cmidrule(l){4-7}
    False & $\ast$ & $-1$ & False & $\ast$ & $\ast$ & $-1$ \\
    True & False & $0$ & True & False & False & $0$ \\
    True & True & $1$ & True & True & False & $1$ \\ \cmidrule(r){1-3}
    &&&True & False & True & $1$ \\
    &&&True & True & True & $2$ \\
     \cmidrule(l){4-7}
    \end{tabular}
\end{table}

\noindent\textbf{Sequential \& Combined Reward:}
We also integrate multiple quality metrics, aiming to train a reward model which models the overall quality of generated tests among several metrics. We expect that a model which optimizes this learned reward will generate high-quality test cases according to multiple metrics.


We developed two strategies for training the reward model.
In the first strategy, called \textbf{Sequential Reward}, we take the policy model trained with Individual Reward (now called the \nth{1} stage model) and re-train it to optimize a different quality metric; we call this the \nth{2} stage model.
Our goal is to preserve the relative simplicity of individual reward training in each stage while optimizing the \nth{2}-stage model for additional properties.


The second strategy, called \textbf{Combined Reward}, combines multiple target reward functions.
For each test, we analyze the properties and assign the rewards for two properties individually, then add the reward values from the two functions.
This yields the composite reward function shown in \Cref{eq:combined-reward}, defined over $k$ quality properties. An example for two properties is shown in \Cref{tab:reward-tables} (right).

\begin{equation}\label{eq:combined-reward}
R_{\text{Comb}}(x_i, y_i) =
\begin{cases}
\sum_{j=1}^{k} R_{\text{P}_j}(x_i, y_i) & \text{if $y_i$ has correct syntax;}\\
-1 & \text{otherwise.}
\end{cases}
\end{equation}

\section{Study Design}

We designed an empirical study with the primary objective of enhancing the quality of test cases generated by Language Model-based approaches through Reinforcement Learning from Static Quality Metrics (RLSQM). Our study aims to address the following research questions:

\begin{itemize}[leftmargin=*]
    \item RQ$_1$: What is the quality of the test cases generated by Codex?
    \item RQ$_2$: Can RLSQM improve Codex to generate high-quality tests?
\end{itemize}

\subsection{RQ$_1$: Assessing Test Case Quality}
\label{sec:rq1-design}



To address RQ$_1$, we generated test cases for a diverse set of focal methods. We generated 100 tests per focal method, then automatically analyzed the test cases, focusing on crucial properties introduced by Peruma et al.~\cite{peruma_distribution_2019}, shown in \Cref{tab:quality-metrics}.

\newenvironment{tinytext}{\begingroup\small}{\endgroup}
\newcommand{\smallboxmargin}{8pt}
\newmdenv[
  innerleftmargin=\smallboxmargin,
  innerrightmargin=\smallboxmargin,
  innertopmargin=\smallboxmargin,
  innerbottommargin=\smallboxmargin,
  outerlinewidth=0.5pt,
  roundcorner=\smallboxmargin
]{smallbox}

\begin{figure}
    \centering
    \caption{Test quality properties. \goodmark{} marks desirable properties and \badmark{} marks test smells.}
    \label{tab:quality-metrics}

    \begin{smallbox}
    \begin{tinytext}
\goodmark{} \textbf{Necessary Test Properties}
\vspace{2pt}
\hrule
\begin{itemize}[leftmargin=*]
    \item \textbf{Correct Syntax:} All test cases should be syntactically valid according to the C\# language specification.
\end{itemize}

\vspace{5pt}

\goodmark{} \textbf{Best Practices}
\vspace{2pt}
\hrule
\begin{itemize}[leftmargin=*]
    \item \textbf{Contains Assertion:} Each test case should contain at least one assertion.
    Tests without assertions can be useful; however, when an assertion is present, a developer can observe its conditions to reason about the purpose of the test.
    \item \textbf{Invokes Focal Method:} Each test case should call the focal method which is specified in the prompt, in order to produce tests relevant to a developer's request.
\end{itemize}

\vspace{5pt}

\goodmark{} \textbf{Documentation}
\vspace{2pt}
\hrule
\begin{itemize}[leftmargin=*]
    \item \textbf{Includes Comment:} The presence of at least one comment, enhancing test case documentation.
    \item \textbf{Has Descriptive Name:} The presence of additional text in the name of the test, which can act as documentation.
    For example, a name stub \texttt{TestAdd} may be enhanced by expanding it to \texttt{TestAdd\_EmptyString\_ReturnsZero}, documenting the input and purpose of the test~\cite{jpreeseBestPracticesWriting2022a}. 
\end{itemize}

\vspace{5pt}

\badmark{} \textbf{Test Smells}~\cite{peruma_distribution_2019}
\vspace{2pt}
\hrule
\begin{itemize}[leftmargin=*]
    \item \textbf{Contains Duplicate Assertion:} Test cases should not contain consecutive identical assertion statements, which may be redundant or inefficient.
    \item \textbf{Contains Conditional Logic or Exception Handling:}
    Test cases containing conditional statements (such as \texttt{if}, \texttt{switch}, \texttt{while}) and exception handling (such as \texttt{try}/\texttt{catch} blocks) are more complex to maintain and reason about.
\end{itemize}
    \end{tinytext}
    \end{smallbox}
\end{figure}

Based on the results, we selected the quality metrics that required the most improvement in the base LLM.
We focus our evaluation on statically-detectable qualitative metrics; dynamic metrics (such as the rate of compilability, passing tests, or coverage) have been studied in other research~\cite{athenatest,a3test,codamosa} and are left to future work.

\subsection{RQ$_2$: Evaluating RLSQM's Effectiveness}
\label{sec:rq2-design}

To investigate RQ$_2$, we compared RLSQM's best-performing training strategy with state-of-the-art code generation models: GPT-4, Base Codex model, and Codex with Supervised Fine-Tuning (SFT) on gold datasets, as proposed in \Cref{sec:sft}.
%
We evaluated all models based on the syntactic correctness and the frequency of quality properties in the tests generated for a held-out set of focal methods.
\temp{We performed several ablations to find the best-performing setting for RLSQM, including training with Individual, Combined, and Sequential rewards.}

\section{Experiments and Results}

\subsection{Experimental settings}
\label{sec:experiments}


\noindent\textbf{Training Procedure:}
We held out 5\% of focal methods for testing and used the remainder for training, sampling the training and test data from different repositories to prevent data leakage. 
%
For all training settings (SFT, RM and PM), we held out 10\% validation data by repository.
We found that validation was critical to avoid mode collapse~\cite{casper2023open} and catastrophic forgetting. During SFT and RM training, we performed early stopping based on the lowest loss on validation data; during PM training, we performed early stopping based on the RM's predicted quality score.
We ensured that all reward models converged to near-0 MSE loss on validation data and all policy models increased the modeled reward $\mathcal{R}$.
Because the RM training data was derived from the Base model's tests, the representation of different test smells was imbalanced; to remedy this bias, we resampled the tests, training the RM on a balanced label distribution.
For complete details of our training procedure, see \Cref{sec:training-detail}.

\noindent\textbf{Base model:}
We generate tests using the OpenAI Codex Cushman model, version \texttt{code-cushman-001}. This version of Codex is a GPT-style model, pretrained on 54 million public repositories using the causal language modeling objective.

\noindent\textbf{Supervised fine-tuning:}
RLSQM was unstable during training when directly applied to the Base model, generating repetitive assertions or comments, or empty tests. To mitigate these behaviors, we initialized RL fine-tuning with the supervised fine-tuned model, following prior literature~\cite{reddit_rlhf, instructgpt}.

Codex generated Documentation properties relatively infrequently (\S~\ref{sec:rq1-results}), resulting in a 50x smaller SFT training set on intersection with the other properties.
The resulting model, denoted \textit{SFT$_{\text{Doc}}$}, improved on Comments and Descriptive Names but regressed on all five of the other properties.
This highlights a limitation of filtering-based SFT -- properties with low frequency or few co-occurrences restricted the size of the golden training dataset.
We trained another SFT model filtered only on Best Practices and Test Smells, which we simply call \textit{SFT}; it regressed on Documentation properties but improved all others, so we used it to initialize all instances of RLSQM.

\noindent\textbf{GPT-4 baseline:}
\renewcommand*{\thefootnote}{\arabic{footnote}}
We compare with GPT-4~\cite{gpt4} (version \texttt{gpt-4-0613}) prompted to generate tests.
Although RLSQM is applicable to any LM, GPT-4 does not support RL fine-tuning.
We generated one candidate test for each focal method due to the increased cost and used the same settings for all generation hyperparameters.
Compared to our simulated prompt, GPT-4 produced more syntactically-correct tests when prompted with natural-language instructions (\texttt{Write a C\# unit test using MSTest for the method \textlangle{}FocalMethod\textrangle{}. Output executable code only.}), followed by the focal code and the prompt hint. We used this prompt to elicit the best performance from GPT-4 and applied adaptive focal context to the focal code to ensure a fair comparison.
%

We trained the models on Azure ML, utilizing virtual machines equipped with 8 NVIDIA A100 GPUs, 96 vCPUs, and 1,924GB RAM. Each GPU had 40GB of HBM2 device memory.
Supervised training took about 16 hours. Reward model training took about 2.5 hours and policy model training took about 4 hours, per run.




\subsection{RQ$_1$: Assessing Test Case Quality}
\label{sec:rq1-results}


\begin{table}[t]
    \centering
    \caption{Codex-generated test quality on the entire corpus of C\# focal methods.
    Frequency denotes the proportion of generated tests containing each smell.
    Test smell priority is based on developer studies~\cite{schvarcbacher_investigating_2019,spadini_investigating_2020}.
    }
    \label{tab:test_properties}
    \begin{tabular}{lrl}
        \toprule
        \textbf{Property} & \textbf{Frequency} & \textbf{Optimization Target}\\
        \midrule
        \goodmark{} \textbf{Necessary Properties} \\
        \midrule
        Correct Syntax     & 83.49\% & - \\
        \midrule
        \goodmark{} \textbf{Best Practices} \\
        \midrule
        Contains Assertion & 69.18\% & Yes \\
        Calls Focal Method & 63.44\% & Yes \\
        \midrule
        \goodmark{} \textbf{Documentation} & & \\
        \midrule
        Descriptive Name  & 12.70\% & Yes \\
        Contains Comment  & 20.71\% & Yes \\
        \midrule
        \badmark{} \textbf{Test Smells} \\
        \midrule
        Duplicate Assertion   & 2.26\% & Yes \\
        Conditional/Exception & 2.31\% & Yes \\
        \midrule
        Assertion Roulette & 18.11\% & No (low-priority) \\
        Magic Number & 17.78\% & No (low-priority) \\
        Sensitive Equality & 2.65\% & No (low-priority) \\
        Redundant Print & 1.40\% & No (infrequent) \\
        Sleepy Test & 0.60\% & No (infrequent) \\
        Empty Test & 0.34\% & No (infrequent) \\
        Mystery Guest & 0.24\% & No (infrequent) \\
        Resource Optimism & 0.01\% & No (infrequent) \\
        
        \bottomrule
    \end{tabular}
    \vspace{-1em}
\end{table}


\definecolor{BaseCodex}{rgb}{0.1946,0.4534,0.6328}
\definecolor{SFT}{rgb}{0.8819,0.5054,0.173}
\definecolor{SFT_Doc}{rgb}{0.2294,0.5706,0.2294}
\definecolor{GPT4}{rgb}{0.7534,0.2387,0.2417}
\definecolor{RLSQM}{rgb}{0.5784,0.4461,0.699}
\newcommand\crule[3][black]{\textcolor{#1}{\rule{#2}{#3}}}
\newcommand{\colorsquare}[1]{\crule[#1]{3mm}{3mm}\,}

\begin{figure*}[t]
    \centering
    \begin{tinytext}
    \begin{tabular}{@{\hskip 1in}c@{\hskip 0.5in}c@{\hskip 0.75in}c@{\hskip 1in}}
    \goodmark{} Necessary/Best Practices &
    \badmark{} Test Smells &
    \goodmark{} Documentation
    \end{tabular}
    \end{tinytext}
    \includegraphics[width=0.8\textwidth]{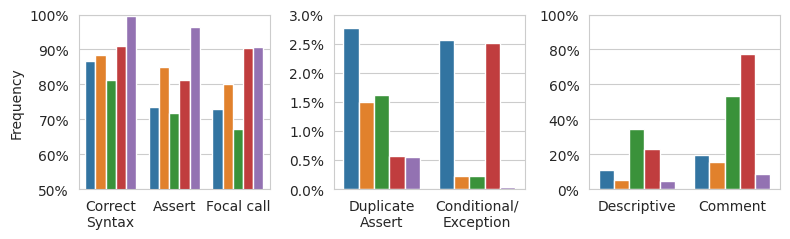}
    \begin{tinytext}
    \begin{tabular}{ccccc}
    \colorsquare{BaseCodex}{Base Codex} &
    \colorsquare{SFT}{SFT} &
    \colorsquare{SFT_Doc}{$\text{SFT}_{\text{Doc}}$} &
    \colorsquare{GPT4}{GPT-4} &
    \colorsquare{RLSQM}{RLSQM}
    \end{tabular}
    \end{tinytext}
    \caption{Comparison between RLSQM's best configuration and baselines. RLSQM outperformed all baselines on all Best Practices and Test Smells, but underperformed on Documentation.
    \protect\footnotemark{}}
    \label{fig:performance-summary}
\end{figure*}

\temp{Codex-generated tests often lacked critical quality properties or contained test smells, both of which contribute to the maintainability and understandability of the test suite.}
\Cref{tab:test_properties} lists the frequencies of the properties we computed on the tests generated by the base model, organized in the sub-categories: Necessary, Best Practices, Documentation, and Test Smells. 
83\% of the base model's completions were syntactically correct, with the remaining 17\% containing syntax errors.
Up to 37\% of generated tests were lacking assertions and 31\% were lacking calls to the focal method, showing room for improvement.
%
Regarding documentation, the base model generated a small percentage of tests containing comments (20.7\%) and descriptive names (12.7\%), marking these as desirable properties for optimization.

\footnotetext{Base model results vary slightly from \Cref{tab:test_properties} because RQ$_1$ evaluates Base Codex on the entire dataset while RQ$_2$ \& RQ$_3$ evaluate models on a held-out subset (\S\ref{sec:experiments}).}

Out of the test smells outlined in Peruma et al.~\cite{peruma_distribution_2019}, we chose to optimize only Duplicate Assertion and Conditional/Exception because we believe that these are most impactful to the developer. Given the large number of experiments required to study the smells in-depth, we had to be selective in the number of properties to optimize.
Six of the test smells not shown were not applicable in our single-test-generation setting.
We regarded the five test smells which occurred in less than 2\% of tests as subpar optimization targets.
Developers considered Conditional Test Logic to have Very High Impact on maintainability in 19.4\% of responses, compared to only 2.1\% for Assertion Roulette and 5.0\% for Magic Number Test (\cite{spadini_investigating_2020} Table 7).
In a prior study, developers considered Sensitive Equality as a false positive or no-action item~\cite{schvarcbacher_investigating_2019}; additionally, this smell can easily be refactored away automatically~\cite{tufano_empirical_2016}.
We consider Duplicate Assertions as likely to confuse the developer by generating redundant code, being affected by side-effects, and obscuring the test's passing conditions.

\begin{table}[htb]
    \centering
    \caption{Codex-generated test quality on C\# focal methods versus Java methods reported by \citet{siddiq2023exploring}.}
    \label{tab:compare-siddiq}
\begin{tabular}{lrrrr}
\toprule
    \textbf{Dataset} & \textbf{UT} & \textbf{DA} & \textbf{CLT\textsuperscript{**}} & \textbf{EH\textsuperscript{**}} \\
\midrule
SF110 (Java) & 21.20\% & 1.40\% & 0.50\% & 20.70\% \\
Ours (C\#) & 26.64\% & 2.77\% & 2.56\% & 2.56\% \\
\bottomrule
\end{tabular}

\small{**We merged CLT and EH into one test smell.}
\end{table}

In \Cref{tab:compare-siddiq}, we also compare our test smells with the results reported on SF110~\cite{siddiq2023exploring}, a set of open-source Java projects similar to our C\# dataset.
While both papers include several test smells not shown, we report the test smells in common on the Codex-2k model.
Siddiq et al. reported similar frequencies for Unknown Test (UT, the inverse of Contains Assertion), Duplicate Assertion (DA), and Conditional Logic Test (CLT).
Exception Handling (EH) occured more frequently in the Java programs, which we believe is due to a difference between the languages; Java programs will not compile if a checked exception is not caught, so the Codex model may tend to catch exceptions inside the test in order to more easily produce a compilable program.

\subsection{RQ$_2$: Effectiveness of RLSQM}
\label{sec:rq2-results}


We ran RLSQM to fine-tune models for each individual quality property, then for combinations of properties.
\Cref{fig:performance-summary} shows the performance of the best-performing RLSQM approach (\ppomodel{\goodmark{} Focal + \badmark{} Conditional/Exception}) compared with baselines.
RLSQM substantially outperformed all baselines on all Necessary/Best Practices and Test Smells, improving upon the Base model by up to 23.2\%, SFT by up to 11.6\%, SFT$_{\text{Doc}}$ by up to 24.7\%, and GPT4 by up to 15.3\%.
Notably, RLSQM produced next to no tests with Conditionals/Exceptions (0.03\%), nearly all its tests were syntactically-correct (99.6\%, compared to 90\% in the best-case for GPT-4).
These results clearly demonstrate that RLSQM is an effective method which substantially improved the quality of tests generated by Codex.
We believe this is enabled by incorporating negative rewards from syntactically-incorrect code, which give policy model direct feedback against generating any syntax errors.

GPT-4, the strongest baseline, produced tests with more Descriptive Names (23\%) and more Comments (78\%) than RLSQM.
We hypothesize that the initial SFT phase of RLSQM, which introduced regressions on Documentation properties, constrained the policy model's ability to enhance these specific properties.
Further research is needed to remove this limitation inherited from the SFT stage.
Since GPT-4 is trained to understand and generate human-like code~\cite{gpt4}, it's not surprising that it produced documentation and avoided consecutive duplicated assertions.
Considering the practical dimension of deployment cost, we estimate that fine-tuned Codex (12B parameters) is several orders of magnitude smaller and costs less than half as much as GPT-4
\footnote{Up-to-date pricing information is no longer available for code-cushman-001 (12B parameters); however, at less than one-tenth the size, we conservatively estimate that RLSQM will cost even less than fine-tuned davinci-002 (175B) at \$0.012/KT (thousand-tokens), which costs less than half of GPT-4 at \$0.03/KT (\url{https://openai.com/pricing}).}, and thus RLSQM can be deployed to generate less-smelly tests at lower cost.
For details of our ablation experiments on optimizing multiple quality properties, see \Cref{sec:rq3-results}.






\section{Related Work}


Previous research on unit test generation has employed evolutionary algorithms, leading to tools such as EvoSuite~\cite{fraser2011evosuite} and Randoop~\cite{pacheco2007randoop}, and several notable machine learning models~\cite{athenatest,a3test,codamosa,toga,siddiq2023exploring,schäfer2023empirical,xie2023chatunitest,yuan2023manual,dakhel2023effective,pan_automated_2021,esnaashari_automation_2021,reddy_quickly_2020,koroglu_reinforcement_2019,koo_pyse_2019}.
The aim of these efforts is generally to resolve compiler errors, produce passing tests from the current code, generate failing tests to expose bugs, or boost code coverage, while we focus on code quality such as test smells, orthogonal qualities which are important for usability and maintainability.
Most closely related, Siddiq et al.~\cite{siddiq2023exploring} demonstrated that ChatGPT and Codex are prone to produce test smells on Python and Java code (see \S~\ref{sec:rq1-results} for a comparison with our findings); however, they do not suggest how to improve the models. In this study, we introduce RLSQM as a method to \textit{enhance} language models based on static quality metrics.
In addition, compared to the existing works, we (1) propose SFT fine-tuning based on filtering gold tests followed by an RL stage, (2) introduce LLM-based reward models, and (3) provide practical guidance about applying RL to LLMs for test generation.

\section{Threats to Validity}

We selected from a large set of 100 projects, including projects from the domains of cloud computing, telecommunications, and game engines, showing that our dataset is indeed diverse. Furthermore, we extracted the public methods which were targeted by existing tests, with the effect of studying the methods the developer intended to test at some point.
This has the added effect of requiring the projects included in the dataset to contain unit tests, which further culls possible toy projects.
Data leakage could occur due to our method of holding out the evaluation dataset. We mitigated this concern by ensuring that the training and test dataset come from different projects, and thus are less likely to share code.

Our experimental results focus on the C\# language due to the availability of our dataset, but we did not evaluate RLSQM on other popular languages like as Java or Python. However, our approach is not restricted to C\#; it can be extended to other languages by re-implementing the simple AST traversals in the Quality Analyzer~\cite{data-package}.
C\#, like Java, is object-oriented, and both languages have a strong emphasis on unit testing. As observed in \Cref{sec:rq1-results}, models produced similar test smell distributions in both languages.
%

Our choice of model and sampling method is inherently non-deterministic, and thus produces different results when run multiple times. We accounted for this variance by generating 100 tests for each method and averaging the results.
Our results also depend on the choice of hyperparameters. We performed manual hyperparameter tuning on the Individual Rewards model to get the best-performing configuration within several settings of the key hyperparameters.



\section{Conclusion}


Reinforcement learning has seen great success in aligning base language models with user intents.
In this paper, we propose Reinforcement Learning from Static Quality Metrics (RLSQM), a fully-automated approach wherein we train a reward model on static analysis-based quality metrics, then optimize a large language model to maximize those rewards.
Our fine-tuning approach involves a basic application of Supervised Fine-tuning, plus an exploration of RL training strategies involving optimizing one or multiple rewards.
We also report how to effectively train models with RLSQM based on our experience and experimental findings.
Our evaluation shows that RLSQM substantially improved the quality of tests generated by Codex, and produced higher-quality tests than GPT-4 at a lower cost.

\newpage

\bibliographystyle{plainnat}
\bibliography{main-short}

\newpage

\clearpage
\begin{appendices}
\crefalias{section}{appendix}

\section{Prompting details}
\label{sec:prompt-details}


\noindent \textbf{Adaptive focal context:} Inspired by Tufano et al.~\cite{athenatest}, we tried a sequence of prompts where each is more concise than the last until the resulting prompt fits within the LLM's focal context window:

\begin{enumerate}
\item Entire text of the focal file, which can include imports, namespaces, and other classes.
\item Focal class code, abbreviating methods other than the focal method to their signatures.
\item Focal class code, with fields and comments removed.
\item Focal class signature and focal method only, discarding all other fields/methods.
\end{enumerate}

Prompts persistently exceeding the permitted length were removed (less than 0.1\% of our dataset).

\noindent \textbf{Sampling:}
We generated tests using nucleus sampling~\cite{Holtzman2020The}.
In order to generate diverse test suites, we set $temperature = 0.7$ and $top\_p = 1.0$ and generated multiple completions for each prompt.

The base model tended to repeat assertions or comments and therefore overrun its completion length, resulting in incomplete and syntactically-incorrect code. To alleviate this problem, we set the frequency penalty to 0.5 to discourage the model from generating tokens which have already been sampled~\cite{frequency_penalty}.

In order to give the maximum context while allowing the model to generate complete and useful tests, we limited completions to 512 tokens (determined empirically) and allowed the context to fill the remaining 1,536 tokens -- this equates to roughly 6,144 characters of context~\cite{openai_tokenizer}.
Our prompts contained about 45 source lines of code (SLOC) on average, measured using \textit{tokei}~\cite{xampprockyTokeiShiJi2023}.
The prompts and generation parameters are in our data package~\cite{data-package}.

\noindent \textbf{Extracting generated tests:} LLMs tended to generate multiple test methods in sequence.
To maintain simplicity, we extracted a single test method from each completion, with the option to generate multiple completions to generate an entire test suite.
To do so, we discarded extra text following the closing brace of the test method, or a second occurrence of the \texttt{[TestMethod]} annotation.

\section{Training procedure}
\label{sec:training-detail}

\noindent \textbf{Training data split:} We split the training data into three equal parts for SFT, RM, and PM training, again separated by repository.
Due to increased computational costs of RL fine-tuning, we randomly sampled 10k prompt/completion pairs for RM training and 1k prompts for PM training.

\noindent \textbf{Validation:} Validation was critical to avoid overfitting in the policy model, such as (1) \textit{mode collapse}~\cite{casper2023open}, where the model learned to generate a narrow band of tests (such as empty tests like \texttt{TestFocalMethod()\{\}}), and (2) \textit{catastrophic forgetting}, where the test generator disregards all quality properties save for the one being optimized.
To avoid these issues, we generated tests for the focal methods in the validation dataset and selected the checkpoint with the best overall test quality, measured by summing the frequency of all best practices and subtracting the frequency of test smells.

\noindent \textbf{Resampling RM training data:} The RM training data, derived from the Base model's generated tests, had an imbalanced frequency of target reward scores and tended to result in biased reward models which impacted PM training. To address this, we resampled the RM training data to maintain a balanced target distribution.
We randomly undersampled~\cite{he2009learning} the majority quality property and negative examples with incorrect syntax; this improved PM convergence and final test quality.

\section{Effects of Individual and Multiple Reward Training}
\label{sec:rq3-results}

To study the effects of training with single rewards versus multiple rewards, we trained RLSQM in an array of configurations using Individual, Sequential, and Combined Rewards and evaluated each configuration on syntactic correctness and overall test quality.
For Individual Rewards, we trained a model for each quality property. The results show that RLSQM was most effective for Best Practices and Test Smells, so we focused on these properties when training multiple rewards.
For Sequential Rewards, we trained a new quality property in the \nth{2} stage.
For Combined Rewards, we evaluated all combinations of 2, 3, or 4 properties.

\begin{figure}[h]
    \centering
    \begin{tabular}{cc}
        \multicolumn{2}{c}{Improvement} \\
        \colorsquare{SFT_Doc} Yes & \colorsquare{GPT4} No \\
    \end{tabular}
    \includegraphics[width=\linewidth]{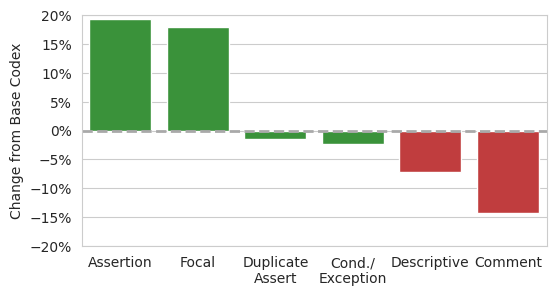}
    \begin{tabular}{@{\hskip 0.5in}c@{\hskip 0.25in}c@{\hskip 0.15in}c}
    \goodmark{} Best Practices &
    \badmark{} Test Smells &
    \goodmark{} Documentation
    \end{tabular}
    \caption{Effects of RLSQM fine-tuning with Individual Rewards. We report change relative to Base Codex.
    The intended outcome is to reduce negative \badmark{} Test Smells and increase positive properties like \goodmark{} Best Practices.
    }
    \label{tab:individual-rewards}
    \vspace{-1em}
\end{figure}

\Cref{tab:individual-rewards} shows the change in the frequency of each quality property after Individual Reward training, compared with the Base model.
RLSQM improved the four Best Practices and Test Smells, notably increasing the frequency of Assertions up to 92.8\% and nearly eradicating Conditional/Exception tests (down to 0.16\% frequency).
By learning from negative examples with incorrect syntax, RL fine-tuning substantially increased the syntactic correctness of their generations, with all RLSQM models achieving 97-99.5\% syntactic correctness. Qualitatively, RL models avoided failure modes such as repeated assertions and comments, which frequently caused the Base and SFT models to generate incomplete tests.

Interestingly, for some properties, the model which optimized one property did not perform as well on that property as a model which optimized another property.
For example, the \ppomodel{\goodmark{} Focal Call} model produced more tests with Assertions than the \ppomodel{\goodmark{} Assertion} model. This may be due to correlation of these properties in the reward model training data, resulting in a RM which learns to reward both properties in a secondary nature.
In our dataset, 34\% of Focal calls occur as arguments to Assertions; therefore, a model that rewards Focal calls may also learn to include the Assertions which commonly wrap them.
Although joint optimization is beneficial in this case and many others, future work can attempt to optimize these properties precisely, without the effect of such correlations.

RLSQM produced fewer Documentation properties than the Base model, which may be due to initialization with SFT weights. We leave addressing this limitation to future work and focus subsequent experiments on optimizing Best Practices and Test Smells.

\label{sec:rq3}

\noindent\textbf{Sequential Reward:}
\Cref{tab:iterative-reward} shows the outcomes of Sequential Reward training.
The properties Assertion and Duplicate Assertion were the most successful, improving in the \nth{2} stage in 5/6 settings.
For 3/4 properties excluding Duplicate Assertions, the \nth{2} stage of RL fine-tuning mostly retained or improved upon the improvements from the \nth{1} stage.
However, compared to the corresponding Individual Reward settings, 6/12 of quality metrics that were optimized in the \nth{2} stage improved or remained nearly unchanged and 6/12 regressed.
While these results show some initial success, further research could seek to more reliably improve the model with Sequential Reward training.

\setlength\fboxsep{2pt}

\begin{table}[t]
    \centering
    \caption{Effects of Sequential Rewards.
    \colorbox{yellow!50}{Yellow cells} mark the properties optimized in the \nth{1} stage; \colorbox{orange!50}{Orange cells} likewise for the \nth{2} stage.
    \textcolor{black!30!green}{Green text} indicates improvement while \textcolor{black!20!red}{red text} indicates regression, relative to the corresponding \nth{1}-stage Individual Reward model.
    }
    \label{tab:iterative-reward}
\begin{tabular}{llcccccc}
\toprule
& \multicolumn{2}{c}{\textbf{\goodmark{} Best Practices}} & \multicolumn{2}{c}{\textbf{\badmark{} Test Smells}} \\ \cmidrule(rl){2-3} \cmidrule(l){4-5}
\textbf{\nth{1} stage $\rightarrow$ \nth{2} stage} & Assert & Focal & \makecell[l]{Dupl.\\Assert} & \makecell[l]{Cond./\\Exc.} \\
\midrule
Best Individual Reward & 92.8\% & 90.9\% & 1.35\% & 0.16\% \\
\midrule
Assert $\rightarrow$ Focal & \colorbox{yellow!50}{\textcolor{black!30!green}{93.5\%}} & \colorbox{orange!50}{\textcolor{black!20!red}{89.9\%}} & \textcolor{black!20!red}{1.42\%} & \textcolor{black!30!green}{0.12\%} \\
Assert $\rightarrow$ \makecell[l]{DA} & \colorbox{yellow!50}{\textcolor{black!30!green}{95.9\%}} & \textcolor{black!20!red}{88.3\%} & \colorbox{orange!50}{\textcolor{black!20!red}{1.44\%}} & \textcolor{black!30!green}{0.09\%} \\
Assert $\rightarrow$ \makecell[l]{CEH} & \colorbox{yellow!50}{\textcolor{black!30!green}{93.1\%}} & \textcolor{black!30!green}{91.6\%} & \textcolor{black!30!green}{0.76\%} & \colorbox{orange!50}{\textcolor{black!20!red}{0.39\%}} \\
\midrule
Focal $\rightarrow$ Assert & \colorbox{orange!50}{\textcolor{black!30!green}{96.7\%}} & \colorbox{yellow!50}{\textcolor{black!20!red}{90.7\%}} & \textcolor{black!20!red}{2.08\%} & \textcolor{black!20!red}{0.31\%} \\
Focal $\rightarrow$ \makecell[l]{DA} & \textcolor{black!30!green}{98.1\%} & \colorbox{yellow!50}{\textcolor{black!30!green}{91.3\%}} & \colorbox{orange!50}{\textcolor{black!30!green}{1.29\%}} & \textcolor{black!20!red}{0.27\%} \\
Focal $\rightarrow$ \makecell[l]{CEH} & \textcolor{black!30!green}{96.4\%} & \colorbox{yellow!50}{\textcolor{black!30!green}{93.4\%}} & \textcolor{black!30!green}{0.94\%} & \colorbox{orange!50}{\textcolor{black!30!green}{0.16\%}} \\
\midrule
DA $\rightarrow$ Assert & \colorbox{orange!50}{\textcolor{black!30!green}{97.4\%}} & \textcolor{black!20!red}{90.4\%} & \colorbox{yellow!50}{\textcolor{black!20!red}{2.78\%}} & \textcolor{black!20!red}{0.29\%} \\
DA $\rightarrow$ Focal & \textcolor{black!30!green}{95.3\%} & \colorbox{orange!50}{\textcolor{black!20!red}{87.8\%}} & \colorbox{yellow!50}{\textcolor{black!20!red}{2.41\%}} & \textcolor{black!20!red}{0.25\%} \\
DA $\rightarrow$ CEH & \textcolor{black!30!green}{94.5\%} & \textcolor{black!20!red}{89.9\%} & \colorbox{yellow!50}{\textcolor{black!20!red}{2.40\%}} & \colorbox{orange!50}{\textcolor{black!20!red}{0.22\%}} \\
\midrule
CEH $\rightarrow$ Assert & \colorbox{orange!50}{\textcolor{black!30!green}{95.8\%}} & \textcolor{black!20!red}{85.6\%} & \textcolor{black!20!red}{1.36\%} & \colorbox{yellow!50}{\textcolor{black!20!red}{0.32\%}} \\
CEH $\rightarrow$ Focal & \textcolor{black!30!green}{96.1\%} & \colorbox{orange!50}{\textcolor{black!20!red}{86.0\%}} & \textcolor{black!20!red}{3.06\%} & \colorbox{yellow!50}{\textcolor{black!30!green}{0.12\%}} \\
CEH $\rightarrow$ DA & \textcolor{black!20!red}{88.7\%} & \textcolor{black!20!red}{80.2\%} & \colorbox{orange!50}{\textcolor{black!30!green}{1.17\%}} & \colorbox{yellow!50}{\textcolor{black!30!green}{0.11\%}} \\
\bottomrule
\end{tabular}
\end{table}

\begin{table}[t]
    \centering
    %
    \caption{Effects of Combined Rewards.
    \colorbox{yellow!50}{Yellow cells} mark the properties under optimization.
    \textcolor{black!30!green}{Green text} indicates improvement while \textcolor{black!20!red}{red text} indicates regression, relative to the corresponding \nth{1}-stage Individual Reward model.
    }
    \label{tab:additive-reward}

\begin{tabular}{lcccccc}
\toprule
& \multicolumn{2}{c}{\textbf{\goodmark{} Best Practices}} & \multicolumn{2}{c}{\textbf{\badmark{} Test Smells}} \\ \cmidrule(rl){2-3} \cmidrule(l){4-5}
\textbf{Model} & Assert & Focal & \makecell[c]{Dupl.\\Assert} & \makecell[c]{Cond./\\Exc.} \\
\midrule
Best Individual Reward & 92.8\% & 90.9\% & 1.35\% & 0.16\% \\
\midrule Assert + Focal & \colorbox{yellow!50}{\textcolor{black!30!green}{93.1\%}} & \colorbox{yellow!50}{\textcolor{black!20!red}{88.4\%}} & \textcolor{black!20!red}{1.60\%} & \textcolor{black!20!red}{0.19\%} \\
Assert + \makecell[l]{DA} & \colorbox{yellow!50}{\textcolor{black!30!green}{94.8\%}} & \textcolor{black!30!green}{91.3\%} & \colorbox{yellow!50}{\textcolor{black!20!red}{1.78\%}} & \textcolor{black!30!green}{0.11\%} \\
Assert + \makecell[l]{CEH} & \colorbox{yellow!50}{\textcolor{black!30!green}{96.8\%}} & \textcolor{black!30!green}{91.5\%} & \textcolor{black!30!green}{1.22\%} & \colorbox{yellow!50}{\textcolor{black!30!green}{0.06\%}} \\
Focal + \makecell[l]{DA} & \textcolor{black!30!green}{96.4\%} & \colorbox{yellow!50}{\textcolor{black!20!red}{90.5\%}} & \colorbox{yellow!50}{\textcolor{black!20!red}{1.46\%}} & \textcolor{black!20!red}{0.20\%} \\
Focal + \makecell[l]{CEH} & \textcolor{black!30!green}{96.5\%} & \colorbox{yellow!50}{\textcolor{black!20!red}{90.8\%}} & \textcolor{black!30!green}{0.55\%} & \colorbox{yellow!50}{\textcolor{black!30!green}{0.03\%}} \\
DA + \makecell[l]{CEH} & \textcolor{black!30!green}{93.1\%} & \textcolor{black!20!red}{87.9\%} & \colorbox{yellow!50}{\textcolor{black!20!red}{1.98\%}} & \colorbox{yellow!50}{\textcolor{black!30!green}{0.15\%}} \\
\midrule Assert + Focal + DA & \colorbox{yellow!50}{\textcolor{black!30!green}{95.3\%}} & \colorbox{yellow!50}{\textcolor{black!20!red}{90.0\%}} & \colorbox{yellow!50}{\textcolor{black!20!red}{1.73\%}} & \textcolor{black!30!green}{0.13\%} \\
Assert + Focal + CEH & \colorbox{yellow!50}{\textcolor{black!20!red}{88.0\%}} & \colorbox{yellow!50}{\textcolor{black!20!red}{76.1\%}} & \textcolor{black!30!green}{0.78\%} & \colorbox{yellow!50}{\textcolor{black!20!red}{0.42\%}} \\
Focal + DA + CEH & \textcolor{black!30!green}{92.9\%} & \colorbox{yellow!50}{\textcolor{black!30!green}{92.8\%}} & \colorbox{yellow!50}{\textcolor{black!30!green}{0.85\%}} & \colorbox{yellow!50}{\textcolor{black!30!green}{0.08\%}} \\
\midrule All 4 properties & \textcolor{black!20!red}{89.8\%} & \textcolor{black!20!red}{84.7\%} & \textcolor{black!30!green}{1.19\%} & \textcolor{black!20!red}{0.21\%} \\
\bottomrule
\end{tabular}
\end{table}

\noindent\textbf{Combined Reward:}
\Cref{tab:additive-reward} shows the results of Combined Reward training.
Two settings improved on all quality properties: \ppomodel{(\goodmark{} Assert + \badmark{} Conditional/Exception)} and \ppomodel{(\goodmark{} Focal + \badmark{} Duplicate assert + \badmark{} Conditional/Exception)}.
Notably, the properties optimized by these models are not intuitively correlated or dependent on each other (compared to e.g. Asserts and Duplicate Asserts, which are certainly interdependent), which may indicate that the Combined Reward setting works most effectively when there are fewer dependencies or potential conflicts between the optimized properties.

However, Combined Reward yielded mixed improvement compared to Individual Reward fine-tuning.
On 2-property optimization, 6/12 optimized properties improved and 6/12 regressed;
on 3-property optimization, 4/9 optimized properties improved while 5/9 regressed;
finally, optimizing All Properties in one reward model performed worse than some 2- or 3-property settings.
The Assert property improved in almost all cases, while the Focal, Duplicate Assert, and Conditional/Exception properties regressed in more than half of the cases.
Along with potential conflicts between quality properties mentioned previously, we hypothesize that this may be because the space of possible rewards grows as more properties are added, ranging from -1 to 4, and thus the policy model got stuck in a local minimum which earned relatively high reward from the reward model, failing to find a balance between exploration and exploitation.
Future work could leverage \textit{reward shaping}~\cite{sutton2018reinforcement} to address this exploration problem.

\noindent\textbf{Summary of RLSQM Training Strategies:}
Overall, RL-finetuned models consistently improved over the Base Codex. The highest-performing model overall was \ppomodel{(\goodmark{} Focal + \badmark{} Conditional/Exception)}, trained with Combined Reward, which substantially improved on all Basic Properties and Test Smells compared to both Codex and GPT-4.

\end{appendices}

\end{document}